\documentclass[11pt]{article}
\usepackage{fullpage,citesort,epsfig,graphics,amsbsy}
\usepackage{psfrag}
\usepackage[normal,small]{caption}
\newcommand{\ignore}[1]{}
 
\usepackage{color}
\definecolor{purple}{rgb}{0.5,0.0,0.5}

\definecolor{green}{rgb}{0.0,0.5,0}
\newcommand{\nn}{\nonumber \\}

\newcommand{\be}{\begin{equation}}
\newcommand{\ee}{\end{equation}}
\newcommand{\beq}{\begin{equation}}
\newcommand{\eeq}{\end{equation}}
\newcommand{\bea}{\begin{eqnarray}}
\newcommand{\eea}{\end{eqnarray}}
\newcommand{\bfs}{\boldsymbol}

\def\to{\rightarrow}

\def\m@th{\mathsurround=0pt }
\def\leftrightarrowfill{$\m@th \mathord\leftarrow 
\mkern-6mu \cleaders\hbox{$\mkern-2mu \mathord- \mkern-2mu$}\hfill
 \mkern-6mu \mathord\rightarrow$}
\def\overleftrightarrow#1{\vbox{\ialign{##\crcr
      \leftrightarrowfill\crcr\noalign{\kern-1pt\nointerlineskip}
      $\hfil\displaystyle{#1}\hfil$\crcr}}}
 
\begin{document}

\begin{titlepage}
\begin{flushright}
 UFIFT-HEP-06-07\\
 Brown-HET-1464\\ 
\end{flushright}
 
\vskip 2.5cm

\begin{center}
\begin{Large}
{\bf String/Flux \hskip-4.2pt Tube Duality on the Lightcone}
\end{Large}

\vskip 2.cm

{\large 
Richard C. Brower,\footnote{E-mail  address: {\tt brower@bu.edu}} 
Chung-I Tan,\footnote{E-mail address: {\tt tan@het.brown.edu}}
and Charles B. Thorn\footnote{E-mail  address: {\tt thorn@phys.ufl.edu}}
}

\vskip 0.5cm
{\it ${}^1$Physics Department, Boston University, Boston, MA 02215}
\vskip 0.5cm
{\it ${}^2$Physics Department, Brown University, Providence, RI 02912}
\vskip 0.5cm
{\it ${}^3$Institute for Fundamental Theory\\
Department of Physics, University of Florida,
Gainesville, FL 32611}


\vskip 1.0cm
\end{center}

\begin{abstract}
\noindent The equivalence of quantum field theory and string theory as
exemplified by the AdS/CFT correspondence is explored from the point of view
of lightcone quantization. On the string side we discuss the lightcone
version of the static string connecting a heavy external quark source to a
heavy external antiquark source, together with small oscillations about the
static string configuration. On the field theory side we analyze the
weak/strong coupling transition in a ladder diagram model of the quark
antiquark system, also from the point of view of the lightcone.  Our results
are completely consistent with those obtained by more standard covariant
methods {in the limit of infinitely massive quarks}.
\end{abstract}


\vfill
\end{titlepage}

\section{Introduction}
\label{sec1}
The postulate of quark confinement in QCD is usually associated
with the idea that, in the theory without dynamical quarks, there
is a mass gap $m_G$ (the lightest glueball mass) and the
gauge field responds to a fixed quark source separated from
a fixed antiquark source by a distance $L\gg m_G^{-1}$ 
by forming some kind of gluonic flux tube between these sources.
The energy of the $q{\bar q}$ system is then expected to grow
with $L$ as $U(L)\simeq T_0 L$. This picture establishes a string-like
object (the flux tube) as a real physical entity in the theory
of strong interactions. Of course, as soon as dynamical quarks enter
the picture, the string can break, making long strings unstable.
But the presence in the hadron spectrum of high spin narrow resonances
on nearly linear Regge trajectories $J\simeq \alpha^\prime M^2$,
with $\alpha^\prime=1/2\pi T_0$, gives strong experimental
evidence that the states of a long string are metastable
and play an important role in the dynamics of the strong interactions.

In the theory with dynamical quarks, 't Hooft's large $N_c$ limit
\cite{thooftlargen} provides another handle, albeit an indirect one,
on quark confinement. It is based on considering a family of 
$SU(N_c)$ gauge groups, QCD being the case $N_c=3$, and
considering the extrapolation $N_c\to\infty$ (with $\lambda=N_c 
\alpha_s/2\pi$
fixed). In this mathematical limit all S-matrix elements
for processes involving only color singlet particles in
the initial and final states are suppressed by powers
of $1/N_c$. In particular the decay rate for any
color singlet particle into any number of color singlet particles
vanishes in the limit. In contrast the decay of
a color singlet into color nonsinglet particles would not be
suppressed. If we define quark confinement as the statement that
all finite mass particles are color singlets, then it follows that
in the limit $N_c\to\infty$ all finite mass hadrons would be stable
noninteracting particles. In particular long strings would be unbreakable 
in this limit. Given the postulate of quark confinement 
we may arguably {\it define} the QCD string as the
large $N_c$ limit of QCD.   There is even hope that the limit is
a quantitatively reasonable approximation to QCD, $N_c=3$.  

One of the big surprises of the duality between superstring theory on
AdS$_5\times$S$_5$ and ${\cal N}=4$ {supersymmetric Yang
  Mills}~\cite{maldacena} is that in the 't Hooft limit a string description
can even be applied to a gauge theory that neither confines nor possesses a
mass gap.  The absence of a mass gap in this case has motivated those
exploring the precise nature of the AdS/CFT correspondence to supply an IR
cutoff by defining the field theory on $S_3$ rather than $R_3$.  Remarkably a
precise correspondence holds in detail in this case as well and this has
enabled many successful quantitative tests of Field/String duality
\cite{bmn}.

However there are some physical situations which provide their
own infrared cutoff (when $N_c=\infty$), and 
these can be safely analyzed on $R_3$.
Among these is the case of fixed quark and antiquark sources
separated by a distance $L$. On the string side of the duality
the absence of infrared difficulties in this system has been implicitly
understood ever since the computation of the large
't Hooft coupling limit of the $q{\bar q}$
potential \cite{maldacenaqqbar} by solving for the static
string configuration that connects the two sources. The point
was further underlined by Callan and G\"uijosa whose
study of small oscillations about this static string configuration
\cite{callang} (see also \cite{Bak}) implies, in the
$\lambda\to\infty$ limit, the existence of many discrete
energy levels above the minimum energy $-c\sqrt{\lambda}/L$ given by the
static solution. From the field theory side, the
existence of stable excited states seems to clash with the absence of
a gap in the field theory; but this puzzle was resolved in
\cite{klebanovmt} where it was shown that at $N_c=\infty$ the
system decouples from all $q{\bar q}+$Glue final states
whose energy is below the ionization threshold $E_{th}=0$.

Thus the large $N_c$ limit of the $q{\bar q}$ system can
be explored in a well-defined way whether or not confinement occurs.
In particular, perturbation theory applied to this
system is infrared safe. In this article we
study various aspects of this system, especially from the
point of view of lightcone quantization. The advantage of the
lightcone description is that it provides an unambiguous
canonical quantization of the system on both sides of Field/String duality.
On the string side, the parametrization of the worldsheet is
completely fixed on the lightcone without the need for
parametrization ghosts. Similarly on the field side the
gauge is completely fixed without the need for Fadeev-Popov
ghosts. Since the quantum evolution on both sides of the duality
is with respect to the same lightcone time, the prospects for a clean and
detailed comparison of the dynamics on both sides of the duality
are especially bright in this formulation.

Indeed, following precisely this line of thought, the lightcone worldsheet
formalism \cite{bardakcit,thorngauge,gudmundssontt} has been developed. This
formalism gives an explicit and concrete lightcone string description of the
sum of the planar diagrams of field perturbation theory in lightcone gauge,
with the goal of understanding field/string duality from the field theory
side.  It is therefore worthwhile to see how the lightcone description deals
with the many aspects of the duality already understood from other starting
points.  On the string side, we test the application of lightcone methods by
showing how the lightcone quantization of string on AdS set up in
\cite{metsaevtt} leads in its semi-classical approximation to the classical
small oscillation problem about a static string stretched between two sources
on the AdS boundary solved by Callan and G\"uijosa. Much of this analysis
goes through with a more general metric than AdS so we only specialize to the
latter at the end.  Of course, at the classical level the lightcone equations
are simply coordinate transformations of the more conventional covariant
equations so it is not surprising to reproduce the solutions in
\cite{callang,Bak}.  It is nonetheless instructive to see how the equations and
their solution flow from the {lightcone starting point which
has no constraints to solve.}  In fact, lightcone quantization is so tightly
formulated that it is hard to describe two completely localized sources. On
the string side, this is because the spatial coordinate $x^-$ is determined
in terms of the other target space coordinates in such a way that fixing the
string ends in those coordinates implies Neumann conditions on $x^-$, i.e.
the string ends move freely in the $z$ direction.  We finesse this difficulty
by fixing the string ends to two separated one-branes parallel to the
$z$-axis {as illustrated in Fig.~\ref{fig:string}}.

On the field theory side, we study the weak coupling
limit of this system by employing the
ladder approximation to the Bethe-Salpeter equation for
fixed separated quark and antiquark sources  
initiated in \cite{ericksonssz}. We discuss this
Feynman gauge model in lightcone coordinates, and we
also construct the corresponding model in lightcone gauge.
We shall see that the lightcone gauge ladder model, 
though different in detail
from the Feynman gauge one, gives qualitatively similar
results.
Finally we turn to a study of the interpolation
between weak and strong coupling within the Feynman gauge
ladder model. This model is only meant as a rough
qualitative guide to the weak/{strong} coupling transition, 
and gives at best a reasonable indication of 
the conformal ${\cal N}=4$ case \cite{klebanovmt}. 
We show that for $\lambda<\lambda_c=1/4$ there are no discrete
levels between the ground state energy and threshold. For
$\lambda>\lambda_c$ an infinite number of levels appear 
accumulating at threshold. These results have
been previously reported in \cite{klebanovmt}, along
with a less detailed derivation of them.
Of course, the ladder 
approximation is strictly valid only at weak coupling,
so the numerical value of this critical coupling should
be taken with a grain of salt. But we think the conformal case
(${\cal N}=4$) should be qualitatively similar. In real
QCD, of course, the 't Hooft coupling depends on the
$q{\bar q}$ separation $\lambda(L)$, so we should then speak
of a critical separation $L_c$. All this suggests that the
confining gluonic flux tube can be usefully thought of as a more or less
conventional multi-gluon bound state, a point of view advocated
in \cite{greensitet}.  

\section{Classical string solutions on the lightcone}
\label{sec3}

The strong 't Hooft coupling limit of the large $N_c$ ${\cal N}=4$
supersymmetric gauge theory is described by the classical IIB superstring on
AdS$_5\times$S$_5$. The fixed quark antiquark system corresponds at strong
coupling to a superstring stretched between two separated points on the AdS
boundary.  Therefore we begin by considering the open string in lightcone
gauge on AdS$_5\times$S$_5$ or a similar background.  We set the worldsheet
fermion fields to zero and focus attention on the bosonic worldsheet
co-ordinates $x^\mu(\sigma,\tau), \phi(\sigma,\tau)$ of the $AdS_5$ factor.
Fixing to lightcone parametrization of the worldsheet, with $x^+ = (x^0 +
x^3)/\sqrt{2} = \tau$ and ${\cal P}^+(\sigma,\tau) = \mbox{1}$, the lightcone
Hamiltonian for {the remaining radial, $\phi$, and transverse, ${\bfs x}
  = (x^1, x^2)$}, coordinates is \cite{metsaevtt}
\begin{eqnarray}
H=p^-={1\over2}\int_{0}^{p^+}{d\sigma} \left[{\bfs{\cal P}}^2
+G^2{\bf x}^{\prime2}
+G(\Pi^2+\phi^{\prime2})\right] \; .
\end{eqnarray}
In fact this is the Hamiltonian for a general warped metric of the form,
\bea
ds^2=G(\phi)dx^\mu dx_\mu + d\phi^2
= G(\phi) (-dt^2 + dx_\perp^2 + dx_3^2) + d\phi^2 \; .
\eea
For the AdS case $G{(\phi)}=e^{\phi/\gamma} = r^2/R^2$ with
$4\gamma^2=R^2T_0=R^2/2\pi\alpha^\prime$.  {So the conformal scaling is
  $\phi\rightarrow \phi + c, x^\mu \rightarrow \exp[-c/2\gamma] x^\mu$.}
However we find it useful to keep $G(\phi)$ arbitrary until the end. As
discussed in reference~\cite{btt}, the more general analysis is useful for a
qualitative understanding of conformally broken QCD-like backgrounds, which
are asymptotically near to $AdS_5$ in the UV ($r \rightarrow 0$) and
terminate beyond some scale in the IR ($r \rightarrow r_0$) to give
confinement.

\begin{figure}
\begin{center}
\psfrag{L}{${\bfs L}$}
\psfrag{x3}{$x^3$}
\psfrag{phi}{$\phi$}
\psfrag{xperp}{${\bfs x}$}
\psfrag{D1}{$D1$}
\psfrag{antiD1}{$\overline{D1}$}
\includegraphics[width=0.40\textwidth]{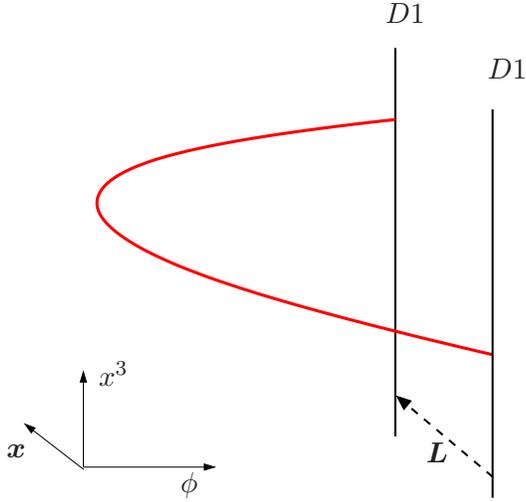}
\caption{String  in  $AdS_5$   fixed at the boundary
as $\phi \rightarrow \infty$ on $D1,\bar D1$ branes aligned
on the $x^3$ axis and separated
by {$L$} in the {${\bfs x} = (x^1,x^2)$} plane.}
\label{fig:string}
\end{center}
\end{figure}

For the purpose of finding classical solutions, it is more convenient to
employ the action implied by this Hamiltonian,
\bea
S=\int d\tau\int_0^{p^+}d\sigma{1\over2}\left[\dot{\bfs x}^2-
G^2{\bfs x}^{\prime2}+{1\over G}\dot{\phi}^2
-G{\phi}^{\prime2}\right]   \; .
\label{action}
\eea
We will be considering solutions in which the ends of the string are
constrained to be on the boundary of AdS. In lightcone quantization, it is
best to preserve $p^+$ conservation. This happens automatically in the
lightcone quantization of the string, which eliminates $x^-$ in favor of the
other worldsheet fields by
\bea
x^{-\prime}=\dot{\bfs x}\cdot{\bfs x}^\prime+\dot\phi\phi^\prime/G
+{\rm fermion~terms} \; ,
\eea
so the Neumann boundary condition $ x^{-\prime}=0$ follows from either
Dirichlet or Neumann boundary conditions on ${\bfs x},\phi$.  However, one
may easily impose Dirichlet conditions on {${\bfs x}, \phi$} at the
string ends so that the string may be stretched a fixed direction ${\bfs L}$
in the ${\bfs x}$ plane at the UV boundary, $\phi \rightarrow \infty$, of
$AdS_5$.  This geometry 
represents infinitely massive $D1, \bar D1$ brane sources
as drawn in Fig.~\ref{fig:string}.  Of course, the equations of motion for
transverse oscillations are not affected by the boundary condition although
the mode discretization does depend on boundary conditions. So the equations
of motion apply equally well to $D0$ (static quarks), $D1$ and $D2$ sources.

\subsection{Static solutions}
The lightcone equations for static string solutions are
\bea
(G^2{\bfs x}^{\prime})^\prime&=&0 \; ,\\
G{\partial G\over\partial\phi}{\bfs x}^{\prime2}
+{1\over2}{\partial G\over\partial\phi}{\phi}^{\prime2}
-(G{\phi}^{\prime})^\prime&=&0 \; .
\eea
These equations imply two integrals of the motion
\bea
{\bfs x}^\prime={\bfs C}/G^2,\qquad {C^2\over G^2}+G\phi^{\prime2}
=2{p^-\over p^+}={M^2\over p^{+2}}\; .
\eea
where ${\bfs C}$ is an
integration constant and $M$ is the total system mass. Note that for static
solutions, the lightcone quantization constraint $x^{-\prime}={\bfs
  x}^\prime\cdot{\bfs{\cal P}} +\phi^\prime\Pi$ implies that there is no
extension in $x^-$.  For an isolated quark, we would also want no extension
in ${\bfs x}$, which would mean ${\bfs C}=0$. For $q{\bar q}$ sources
separated by a distance $L$ we have 
\be
L=\left|\int^{p^+}_0 d\sigma {\bfs
    x}^\prime\right| = C\int^{p^+}_0 {d\sigma\over G^2} \; ,
\ee
which together with the integrals of the motion determine the system
mass{, $M = \sqrt{2 p^+ p^-}$}. For the AdS case $G=e^{\phi/\gamma}$ the
static string's mass was found to be \cite{metsaevtt}
\begin{eqnarray}
M&=& 4\gamma\sqrt{G_{\rm max}}-{4\gamma^2\over L}
{(2\pi)^3\over\Gamma(1/4)^4} \; .
\label{eq:mass}
\end{eqnarray}
Which agrees with the known result \cite{maldacenaqqbar}, after
we recall that 
$4\gamma^2=R^2T_0=R^2/2\pi\alpha^\prime=\sqrt{\alpha_s N_c/\pi}$.
The first divergent term is just twice the isolated quark mass,
so the second finite term is the predicted interaction energy 
between {infinitely massive} quark and antiquark.
\subsection{Small transverse oscillations}
On the lightcone the ends of a static string stretched
between two fixed points in transverse space automatically lie
at the same longitudinal coordinate because $x^{-\prime}=0$. 
However, to calculate
small oscillations about this static configuration on the lightcone,
it is simplest to retain fixed Dirichlet conditions only on the
transverse coordinates of the string ends, but allow the 
ends to move freely in the longitudinal direction, in the manner
suggested in \cite{rozowskyt}. Fixing the
longitudinal coordinates of the ends would impose a nonlinear constraint
on the transverse oscillations which is awkward to implement, and would
defeat the main purpose of using lightcone in the first place.
Thus for lightcone quantization 
we shall instead consider a string stretched between two fixed
1-branes parallel to each other and to the $z$ axis, and separated
from each other by the distance $L$. As just explained the 
static solution for this physical situation is identical
to that quoted in the previous subsection. 

To study small oscillations, we replace ${\bfs x}, \phi$ by 
{${\bfs x(\sigma)}_c+{\bfs
  x}, \; \phi_c(\sigma)+\phi$} and expand the action to quadratic 
order in ${\bfs
  x}, \phi$. The classical equations of motion for ${\bfs x}_c,\phi_c$
guarantee that the linear terms in fluctuations vanish and we get
\bea
S= S_c + \int d\tau\int_0^{p^+}d\sigma{1\over2}\bigg[\dot{\bfs x}^2 &-&
G_c^2{\bfs x}^{\prime2}+\dot{\phi}^2G_c^{-1}
-G_c{\phi}^{\prime2}
-\left((G_cG_c^{\prime\prime}+ G_c^{\prime2}){\bfs x}_c^{\prime2}
+{1\over2}{\phi}_c^{\prime2}G_c^{\prime\prime}\right){\phi^2}
\nonumber\\
&-&4G_cG_c^\prime\phi{\bfs x}^{\prime}\cdot{\bfs x}^{\prime}_c
-{2\phi_c^\prime}G_c^\prime\phi\phi^\prime\bigg] \; ,
\eea
where {$G_c(\sigma)= G(\phi_c(\sigma))$ and $G_c^\prime(\sigma)=
  \partial G/\partial\phi$ evaluated at $\phi = \phi_c(\sigma)$,
 etc.} In the special AdS limit,
\bea
S \to S_c  +\int d\tau\int_0^{p^+}d\sigma{1\over2}\bigg[\dot{\bfs x}^2 &-&
e^{2\phi_c/\gamma}{\bfs x}^{\prime2}+\dot{\phi}^2e^{-\phi_c/\gamma}
-{\phi}^{\prime2}e^{\phi_c/\gamma}
-2\left(e^{2\phi_c/\gamma}{\bfs x}_c^{\prime2}
+{1\over4}{\phi}_c^{\prime2}e^{\phi_c/\gamma}\right){\phi^2\over\gamma^2}
\nonumber\\
&-&{4\phi\over\gamma}
e^{2\phi_c/\gamma}{\bfs x}^{\prime}\cdot{\bfs x}^{\prime}_c-{2\phi_c^\prime
\over\gamma}
e^{\phi_c/\gamma}\phi\phi^\prime\bigg]\; .
\eea
We see that the oscillations in ${\bfs x}$ perpendicular to ${\bfs x}_c$
decouple from the $\phi$ oscillations, whereas those parallel to
${\bfs x}_c$ have a coupling to the $\phi$ oscillations. 
{In this sub-section, we 
restrict attention to transverse oscillations  where 
$ \bfs x\cdot {\bfs x}_c  =0$. Two other independent 
modes will be considered in the following  sub-sections.}  
Letting $\omega$ be the oscillation frequency, we see that the
transverse oscillations are governed by the ordinary differential
equation
\bea
\omega^2 f+\partial_\sigma(G_c^2\partial_\sigma f)&=&0\nn
f|_{\sigma=0, p^+}&=&0 \; .
\label{eq:eomtrans}
\eea

\begin{figure}
\begin{center}
\psfrag{\xi}{$\xi$}
\psfrag{dxi}{$\Delta \xi/\xi$}
\includegraphics[width = 0.45\textwidth]{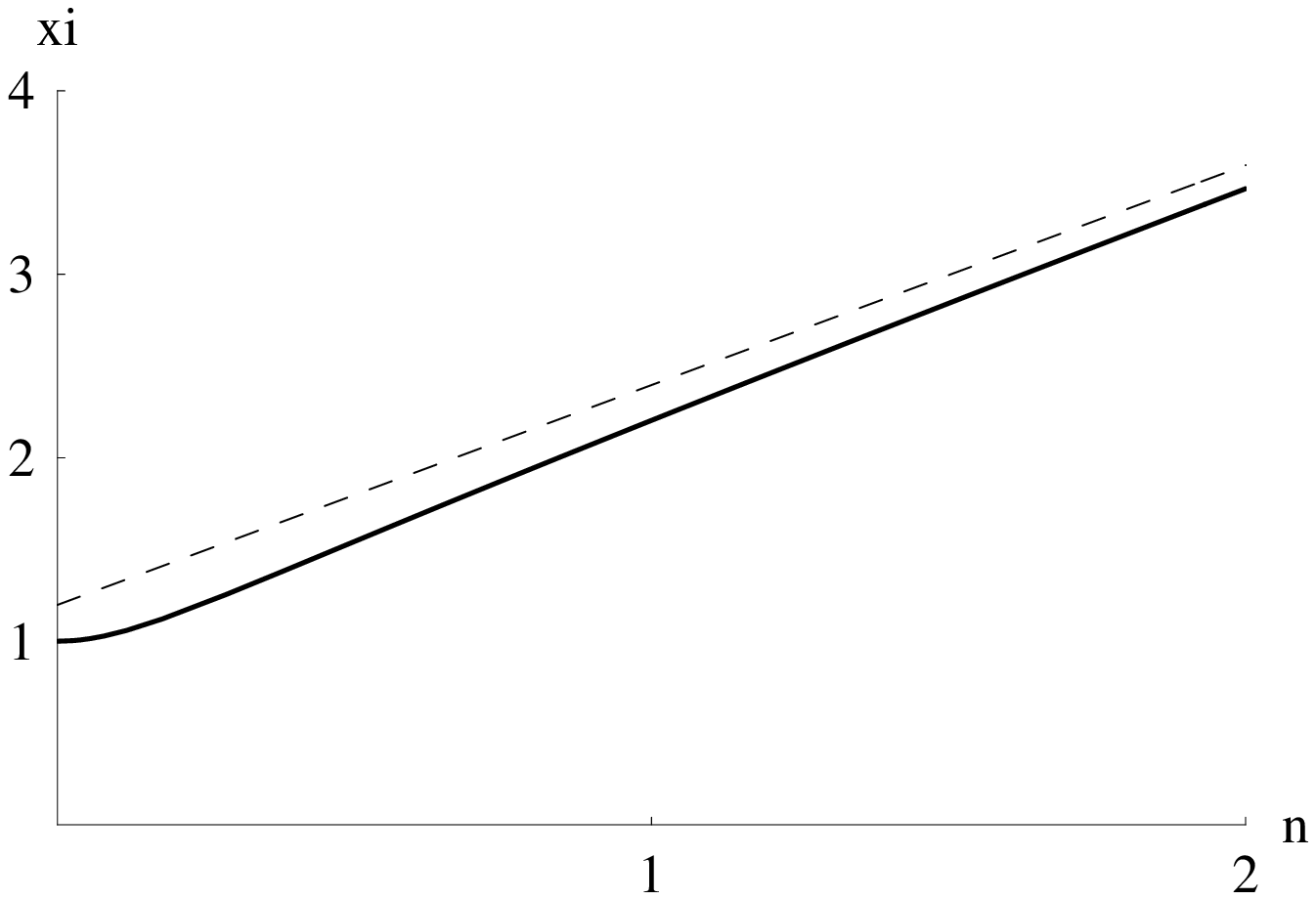} 
\includegraphics[width = 0.45\textwidth]{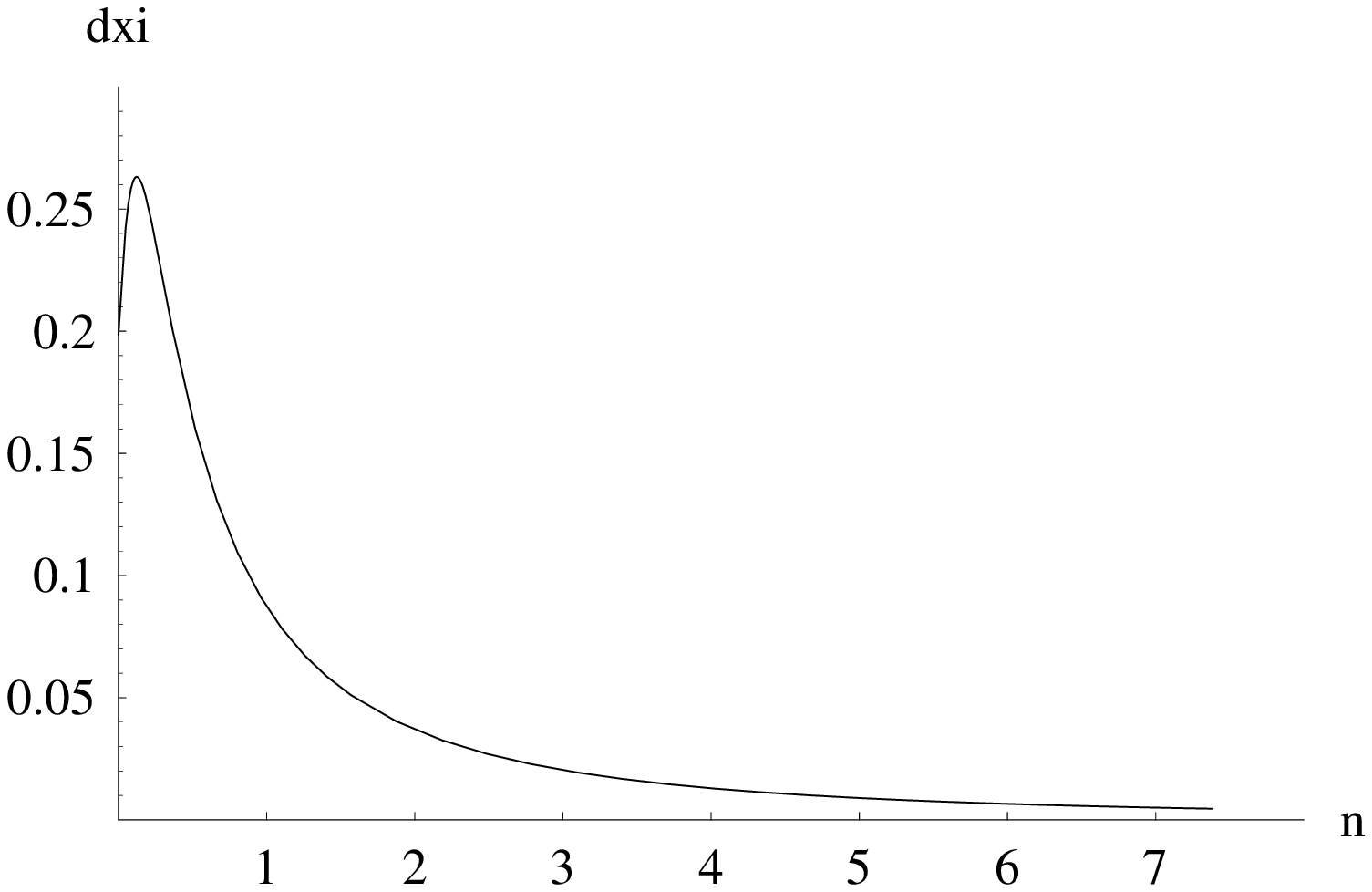}
\caption{{On the left, 
the numerical solution to eigenvalues $\xi_n$ 
for transverse string fluctuations
in $AdS_5$ (solid line) versus the large n asymptotic expansion $\xi_n =
(n+1)(2 \pi)^{3/2}/\Gamma(1/4)^2 + O(1/n)$ (dotted line) and on the right
the fractional error in the asymptotic expression for $\xi_n$.} }
\label{fig:xi}
\end{center}
\end{figure}

We now specialize to the AdS case for which
$G_c=r^2_c/R^2=e^{\phi_c/\gamma}$.
In order to simplify notation, we will also set $R=1$ in what follows.   
Since the static solution $r_c(\sigma)$ is piecewise
monotonic, it is possible to change variables from $\sigma$ to $r_c$ over
each monotonic interval. Recall that
\bea
r_c^{\prime2}={C_\perp^2\over 4 \gamma^2}\left({1\over r_{min}^4}
-{1\over r^4_{c}}\right) \; .
\eea
The fact that $\sigma$ measures $p^+$ implies that
\bea
{C_\perp\over\gamma}={2 r^{3}_{min}\over p^+}
\int_1^{r^2_{max}/r^2_{min}}{d\chi\over\sqrt{\chi-1/\chi}}
\simeq{4r^2_{min}{r_{max}}\over p^+} \; .
\eea 
Then we find, dispensing with the $c$ subscript,
\bea
 {dr \over d\sigma}&=&\pm{2 r^2_{min} r_{max}\over p^+}
\sqrt{{ 1\over r_{min}^4}-{1\over r^4}}\\
\omega^2 f&+&\left({2r^2_{min}{r_{max}}\over p^+}\right)^2
\sqrt{{1\over r_{min}^4}-{1\over r^4}}
\partial_r \left(r^4\sqrt{{1\over r_{min}^4}-{1\over r^4}}
\partial_ r  f\right)=0\; .
\eea
Another useful coordinate is  $z\equiv R^2/r$ so {with $R=1$}, 
we also write the small oscillation equation as,  
\bea
\omega^2 f+\left({2 r_{max}}\over p^+\right)^2
\left[(1-z^4 r_{min}^4)
\partial^2_z f - {2\over z}\partial_z f\right]=0 \; .
\eea
Now put $\omega=2\xi{r_{min}r_{max}}/p^+$
and ${\hat z}= r_{min} z$ to reduce the equation
to that solved by Callan and G\"uijosa\cite{callang},
\bea
\xi^2 f+(1-{\hat z}^4)\partial^2_{\hat z} f - {2\over {\hat z}}
\partial_{\hat z} f=0 \; .
\eea
This equation can be brought into a more familiar form,
\bea
\frac { d^2 F(q) }{dq^2  }+ \frac{\xi^2(\xi^4-1)}{4} F(q)=0 \; ,
\eea
where $F = f/ \sqrt {1+\xi^2 {\hat z}^2}$, and 
{$q = \pm 2 \int_{\hat z}^1 d t\; t^2/[(1+\xi^2t^2){\sqrt{1-t^4}}]$}.
With Dirichlet boundary conditions and $r_{max}=\infty$ the 
exact eigenfrequency equation becomes
\bea
\xi_n\sqrt{\xi_n^4-1}\int_0^1{t^2dt\over[1+\xi_n^2t^2]
\sqrt{1-t^4}}={n\pi\over2},\qquad n=1,2,\cdots 
\eea
The integral here is an elliptic integral and the equation must be
solved numerically. However for large frequency it becomes elementary,
\beq
\xi_n \simeq {(n+1)\pi\over2}\left[\int_0^1{dt\over\sqrt{1-t^4}}\right]^{-1} 
=(n+1){({2\pi})^{3/2}\over\Gamma(1/4)^2}  \; .
\eeq
The next correction is $O(1/n)$. In fact it turns out that the exact $n$
versus frequency curve is practically linear for $n\geq1$ {(see
Fig.~\ref{fig:xi})}.  This asymptotic form is the asymptote for the exact
curve, the exact frequency being roughly 10\% below the asymptote for $n=1$.

We must now relate the $\xi$'s to actual physical energy levels.  Since on
the lightcone the Hamiltonian is $p^-$, the $\omega$'s represent the
excitations in this variable. {To match with the semi-classical solution
  in temporal quantization given below we must compare the finite excitation
  energies in the limit to infinitely massive source ($r_{max} \rightarrow
  \infty$).}   In the previous section in
{Eq.~\ref{eq:mass}}, we found that the {$p^- = M^2/2 p^+$} value of
the static string solution was
\bea
p^-_c&=&{1\over2p^+}\left[4\gamma {r_{max}}-{4\gamma^2\over L}
{(2\pi)^3\over\Gamma(1/4)^4}\right]^2 \nn
&\simeq&{1\over2p^+}\left[(4\gamma r_{max})^2
-8\gamma r_{max}{4\gamma^2\over L}
{(2\pi)^3\over\Gamma(1/4)^4}+O(1)\right] \; ,
\eea
as $r_{max}\to\infty$. To this we add the excitation energy
\bea
\Delta_{\{N_i\}}=\sum_n N_{n}\omega_n\simeq{4 {r_{max} r_{min}}
\over2p^+}\sum_n N_{n}\xi_n={8\gamma{r_{max}}
\over2p^+L}{({2\pi})^{3/2}\over\Gamma(1/4)^2}\sum_n N_{n}\xi_n  \; .
\eea
Thus we find (restoring the $R$ dependence by $r_{max}\rightarrow r_{max}/R$)
\bea
p^-_{\{N_i\}}
&\simeq&{1\over2p^+}\left[\left(4\gamma{r_{max}\over R}\right)^2
+8\gamma {r_{max}\over R}\left(-{4\gamma^2\over L}
{(2\pi)^3\over\Gamma(1/4)^4}+{1\over L}
{({2\pi})^{3/2}\over\Gamma(1/4)^2}\sum_n N_{n}\xi_n\right)
+O(1)\right]   \; , \nonumber\\
E^{CM}_{\{N_i\}}&=&\sqrt{2p^+p^-_{\{N_i\}}}
\simeq4\gamma{r_{max}\over R}-{4\gamma^2\over L}
{(2\pi)^3\over\Gamma(1/4)^4}+{1\over L}
{({2\pi})^{3/2}\over\Gamma(1/4)^2}\sum_n N_{n}\xi_n  \; .
\label{pminusenergy}
\eea
{As we will now show this excitation spectrum for $M \rightarrow \infty$
  is in exact agreement with the semi-classical approximation using temporal
  coordinates $x^0 = \tau$ for all $L$. 
  Moreover, it is interesting that for flat
  space, where the lightcone action is quadratic and therefore the 
  semi-classical solution to the stretched string is exact~\cite{Arvis:1983fp},
  the agreement with temporal quantization also requires taking the $L
  \rightarrow \infty$ limit. In this limit the leading excitation energy is the
  $1/L$ universal conformal ``L\"uscher term''.  Interestingly in the
  present case of a truly conformal $AdS_5$ theory only $M \rightarrow
  \infty$ is needed for agreement between the  leading lightcone and temporal  semi-classical
  approximations.}

\subsection{Comparison to temporal quantization}
{We have so far identified only one transverse mode in our lightcone
  analysis.} We consider here the small oscillation problem using {
the temporal parameters where the evolution parameter is
target space time, $t=x^0=\tau$.}
{In this parametrization, 
it will be obvious that there are two degenerate modes
  transverse to a stretched string, and another mode for ``longitudinal"
  oscillation. We shall return in the next sub-section to see how these two
  additional modes manifest themselves in a lightcone approach.}

To do this most systematically, we use the phase space
version of Hamilton's principle:
\bea
{\cal L}= {\dot{\vec x}}\cdot{\vec{\cal P}}+{\dot \phi}\Pi-{\cal P}^0
-{\lambda\over2}\left({\vec{\cal P}}^2+G^2(\phi){\vec x}^{\prime2}
-{\cal P}^{02}
+G(\phi)(\Pi^2+\phi^{\prime2})\right)-\mu({\vec x}\cdot{\vec{\cal P}}
+\phi^\prime\Pi) \; ,
\eea
where $\lambda,\mu$ are Lagrange multipliers implementing the constraints
associated with parameterization invariance. The lost equation from setting
$x^0=\tau$ is ${\dot{\cal P}}^0-(\mu{\cal P}^0)^\prime=0$.  Integrating out
${\cal P}^0$ gives ${\cal P}^0=1/\lambda$. With a time dependent
reparametrization of $\sigma$ we can set $\mu=0$, and with a further time
independent one we can set $\lambda=1$ ($\mu=0$ and ${\cal P}^0=1/\lambda$
imply ${\dot\lambda}=0$).  Doing this fixes the scale of $\sigma$ whose
length is no longer arbitrary: $0<\sigma<E$ where $E$ is the total energy of
the motion.  Also we must remember the constraints which will no longer be
implied by Hamilton's principle:
\bea
{\cal L}&=& {\dot{\vec x}}\cdot{\vec{\cal P}}+{\dot \phi}\Pi
-{1\over2}\left(1+{\vec{\cal P}}^2+G^2(\phi){\vec x}^{\prime2}
+G(\phi)(\Pi^2+\phi^{\prime2})\right)\\
1&=&{\vec{\cal P}}^2+G^2(\phi){\vec x}^{\prime2}+G(\phi)(\Pi^2+\phi^{\prime2})
\\
0&=&{\vec x}\cdot{\vec{\cal P}}+\phi^\prime\Pi  \; .
\eea
At this point we can integrate out momenta to return to the coordinate space
Lagrangian: 
\bea
{\cal L}&=& {1\over2}\left({\dot{\vec x}}^2+{{\dot \phi}^2\over G(\phi)}
-1-G^2(\phi){\vec x}^{\prime2}-G(\phi)\phi^{\prime2})\right)\\
1&=&{\dot{\vec x}}^2+G^2(\phi){\vec x}^{\prime2}+{{\dot\phi}^2\over G(\phi)}
+G(\phi)\phi^{\prime2}\label{first}
\\
0&=&{\vec x}\cdot{\dot {\vec{x}}}+{\phi^\prime{\dot \phi}\over G(\phi)} \; .
\label{second}
\eea 
{The need for these side constraints, which are absent on the lightcone,
  is the main obstacle to quantization in this parametrization.} Classically
  there is of course no problem with them.  If the second constraint holds,
the right side of the first constraint is a constant of the motion following
from the equations of motion for ${\vec x},\phi$. It can then be used in
place of one of the equations of motion. Thus to find all the classical
solutions we only need solve all but one of the equations, 
\bea
{\ddot {\vec x}}-(G^2{\vec x}^\prime)^\prime&=&0\\
{{\ddot\phi}}-G^2\phi^{\prime\prime} +G^2{\partial G\over\partial\phi}{\vec
  x}^{\prime2}-{1\over2} {\partial
  G\over\partial\phi}\left(G\phi^{\prime2}+{{\dot \phi}^2 \over G}\right)&=&0
\; ,
\eea 
together with the two constraints.  For small oscillations around a
static solution it will be useful to use the constraint to eliminate ${\vec
  x}^{\prime2}$ from the $\phi$ equation: 
\bea
{{\ddot\phi}}-G^2\phi^{\prime\prime} -{3\over2}G{\partial
  G\over\partial\phi}\phi^{\prime2} +{\partial G\over\partial\phi} -{\partial
  G\over\partial\phi}\left({\dot{\vec x}}^{\prime2}
  +{3\over2}{{\dot\phi}^2\over G}\right)&=&0\\
{{\ddot\phi}} +{1\over
  2G\phi^\prime}\left[G^2(1-G\phi^{\prime2})\right]^\prime -{\partial
  G\over\partial\phi}\left({\dot{\vec x}}^{\prime2}
  +{3\over2}{{\dot\phi}^2\over G}\right)&=&0 \; .
\label{eomphi}
\eea
For small oscillations about a static solution the last
term is second order in fluctuations and can be dropped.
Without the constraints,
the equations for $\phi, {\vec x}$ are identical in form
to the lightcone equations for $\phi, {\bfs x}$. But we must
bear in mind that here $\tau$ is ordinary time rather than
lightcone time and $\sigma$ is a measure of ordinary energy rather
than of $p^+$.

For static solutions, the second constraint is 
automatically satisfied and the equations for ${\vec x}$
are immediately integrable, so it is convenient to
drop the $\phi$ equation of motion
and obtain
\bea
{\vec x}^\prime={\vec C\over G^2},
\qquad \phi^{\prime2}={1\over G}-{C^2\over G^3} \; , 
\eea
{where $\vec C$ is again a constant vector.}  For fixed ends separated by the vector ${\vec L}$ we have
\bea
E&=&\int_0^E d\sigma = 2\int_{\phi_{min}}^{\phi_{max}}
d\phi {\sqrt{G}\over\sqrt{1-C^2/G^2}}\\
{\vec L}&=& \int_0^E d\sigma {\vec C\over G^2}=2{\vec C}
\int_{\phi_{min}}^{\phi_{max}}
d\phi {\sqrt{G}\over G^2\sqrt{1-C^2/G^2}} \; ,
 \eea
where the constant $C$ can be fixed by the condition $\phi'=0$, leading to
$C=G(\phi_{min})$.  For the AdS case ${G(\phi)}=e^{\phi/\gamma}\equiv
r^2$, so $d\phi= 2 \gamma d r/ r $, $r^2_{min}=C$ and we obtain
\bea
E&=& 4 \gamma\int_{r_{min}}^{r_{max}}
dr  {r^2\over\sqrt{r^4-r_{min}^4}}\\
|{\vec L}|&=& 4 \gamma r^2_{min}\int_{r_{min}}^{r_{max}}
dr  {1\over r^2\sqrt{r^4-r_{min}^4}} \; .
\eea
\arraycolsep 2pt
We see that $L, E$ coincide with the lightcone evaluations 
of $L, \sqrt{2p^+p^-}$ as they must.

Now we come to small oscillations about the static solution ${\vec x}_c,
\phi_c$. From now on call the fluctuations about these solutions ${\vec x},
\phi$.  First take the case of oscillations, $\vec x_\perp$, transverse to
${\vec L}$. In this case the ${\vec x_\perp}$ fluctuations enter the
constraints and the equation for $\phi$ quadratically, so we can consistently
set $\phi=0$, and seek solutions of
\bea
{\ddot {\vec x_\perp}}-(G^2(\phi_c){\vec x_\perp}^\prime)^\prime&=&0 \; .
\eea 
As we have already mentioned this 
differential equation is identical in form
to the {lightcone} equation for ${\bfs x}$. However, there are differences
in the passage to the Callan-G\"uijosa form. Namely in the relation 
between $\omega$ and $\xi$. For ordinary time we find
$\omega_{temp}=\xi r_{min}/2\gamma$ compared to the
lightcone relation $\omega_{LC}=2\xi r_{min}r_{max}/p^+$.
The different prefactors here precisely account for the
fact that $\hbar\omega_{LC}$ gives the semi-classical 
excitation energy of lightcone energy $p^-$, whereas $\hbar\omega_{temp}$
gives the excitation energy of ordinary energy $E$. This is easily
seen by comparing the two lines of (\ref{pminusenergy}). Here we see that
there are two modes of transverse oscillation with identical frequencies.
On the lightcone, in contrast, we only see one manifest transverse mode. 
We return to the other transverse lightcone mode, related
to fluctuations in $x^-$, at the end of {the next}  subsection.

Next we consider oscillations,  $ x_\parallel$, {parallel} to ${\vec L}$.
For definiteness, take ${\vec L}=L{\hat x}$, where $\hat x$ is also one of
the lightcone transverse directions.  In temporal quantization 
the equations to
solve couple the $x_\parallel$ coordinate with $\phi$. 
The linearized constraints become
\bea
G^2{(\phi_c)}({x_c^{\prime2}+2x_c^\prime x^\prime_\parallel})
+G{(\phi_c)}(\phi_c^\prime+\phi^\prime)^{2}&=&1\label{firstlin}
\\
{x_c^\prime}{\dot x_\parallel}+{\phi_c^\prime{\dot \phi}\over G(\phi_c)}
&=&0\label{secondlin} \; ,
\eea
which then serve to express ${\dot x}_\parallel, x^\prime_\parallel$ as explicit functions
of $\phi$.
Thus the longitudinal small oscillations are completely controlled by
the linearized approximation of (\ref{eomphi}),
\bea
{\ddot \phi}+{1\over G_c\phi_c^\prime}\left[-G_c^3\phi_c^\prime\phi^\prime
+G_c{\partial {G_c}\over\partial\phi}_c\left(1-{3\over2}G_c\phi_c^{\prime2}\right)
\phi\right]^\prime&=&0 \; ,
\eea
where {$G_c(\sigma) = G(\phi_c(\sigma))$.}

It is useful to change variable from $\sigma$ to $x_c$, which measures the
length of the stretched string, i.e., using $ x_c'= C/G_c^2$ and $-L/2 \leq
x_c\leq L/2$. One finds that the above linearized equation can be written in
a generic form\cite{btt}
\bea
 (\partial_t^2- v^2(x_c)\partial^2_{x_c} + m^2(x_c) )\Psi(t,x_c) = 0 \; ,
\eea
where 
\bea
m^2(x_c)= (G_0^2/G_c^2)[G_c^{''}- (3/2) {G_c'}^2/G_c]\; ,
\eea
and $v^2= (G_0^2/G_c^2) $.  Note that the wave function has been rescaled,
$\Psi =G_c^{1/2}\phi$, and that the derivatives are taken respect to $\phi$, $G'_c
= \partial G/\partial \phi$, $G''_c = \partial^2 G/\partial \phi^2$ evaluated
at $\phi = \phi_c$. We have also introduced $G_0\equiv C= G(\phi_{min})$ to
rationalize the notation. In deriving this result, we have made use of the
fact that
\bea
\phi_c'= G_c^{-3/2}(G_c^2-G_0^2)^{1/2}\; .
\eea
This longitudinal mode has been identified previously 
in Ref. \cite{btt} as the ``radion" mode. 
In particular, for a metric deformation with confinement, 
this mode is massive, with the mass scale 
set by the glueball mass\footnote{We thank Igor Klebanov for
drawing our attention to a typo in Eq. (23) in Ref. \cite{btt} 
in the expression for  the mass. There $G(z)$ has 
a different definition by expressing the metric as
 $ds^2 =\frac{R^2}{z^2}[ dx^2 + G^2(z) dz^2]$, which
should have led to the expression
$$m^2(x_c)= -\frac{1}{z_{max}^4} [z^3\frac{d}{dz} 
\frac{1}{G^2} + \frac{2z^2}{G^2}]_{z=z_c}. $$}.
Here, for pure $AdS$, the frequency scales as $1/L$.

\subsection{{Radion and second transverse modes on the lightcone}}

In lightcone quantization, we have the equation of motion for $x$, 
and the equation of motion for
$\phi$. But there are no further constraints: instead the right side
of (\ref{first}) is just $2{\dot x}^-$ and the right side of
(\ref{second}) is $x^{-\prime}$: 
\bea
{\dot x}^-&=&{1\over2}\left({\dot{\bfs x}}^2+G^2(\phi){\bfs x}^{\prime2}
+{{\dot\phi}^2\over G(\phi)}
+G(\phi)\phi^{\prime2}\right) \; ,
\label{xminusdot}
\\
x^{-\prime}&=&{\bfs x}^\prime\cdot{\dot {\bfs{x}}}
+{\phi^\prime{\dot \phi}\over G(\phi)} \; .
\label{xminusprime}
\eea 
The integrability condition for this pair of equations, that the $\sigma$
derivative of the first equals the $\tau$ derivative of the second, follows
from the equations of motion for ${\bfs x},\phi$.  But for strictly
longitudinal classical motion we do want $x^{-\prime}=0$, in which case
(\ref{second}) will also be imposed on the lightcone. Then the right side of
(\ref{xminusdot}) is a constant of the motion fixed to be $2p^-/p^+$ by the
fact that ${\dot x^{-}}$ is the density of {$p^-$} momentum.  The
equations for longitudinal oscillations on the lightcone can be made
identical to those in temporal quantization by scaling lightcone
$\sigma,\tau$ by a common factor $\sqrt{p^+/2p^-}$. This implies that the
oscillation frequencies in the two approaches will be related by
$\omega_{LC}=\omega_{ord} \sqrt{2p^-/p^+}$, {in the limit of infinitely
  massive sources.}  {In turn this}
implies that the total energy calculated in temporal quantization will agree
with $\sqrt{2p^+p^-}$ calculated on the lightcone:
\bea
\sqrt{2p^+(p^-_0+\hbar\omega_{LC})}\simeq \sqrt{2p^+p^-_0}\left
(1+\hbar{\omega_{LC}\over 2p_0^-}\right)=\sqrt{2p^+p^-_0}+\hbar\omega_{ord}
\; .
\eea
Here $p^-_0$ is the lightcone energy of the static solution.  Thus as far as
longitudinal oscillations between infinitely massive sources are concerned
the two approaches agree.

Finally, we consider the second transverse oscillation from the point of view
of lightcone.  Then we look for solutions in which fluctuations in both
transverse lightcone coordinates vanish. Then the fluctuations are completely
described by fluctuations in $\phi$. The oscillation frequencies could be
determined by analyzing the $\phi$ equation of motion.  However, 
it is quicker to find the equation that $x^-$ satisfies because it is a
function of ${\bfs x}, \phi$. Indeed for any ${\bfs x}, \phi$ satisfying
their respective equations it is straightforward to compute 
\bea 
{\ddot  x}^--(G^2 x^{-\prime})^\prime=0  \; .
\eea
This is the same differential equation
satisfied by ${\bfs x}$.  For the static solution $x^{-\prime}_c=0$, so the
small oscillation approximation implies 
\bea 
{\ddot x}^--(G_c^2x^{-\prime})^\prime=0 \; ,
\eea 
the same equation as oscillations in ${\bfs x}$
perpendicular to ${\bfs L}$.  The only difference is that the way we have set
up boundary conditions on the lightcone, Dirichlet conditions will not apply
to $x^-$.  Indeed, we have Dirichlet conditions on ${\bfs x},\phi$. This
clearly implies Neumann conditions on $x^-$. Thus with the lightcone setup
the eigenfrequencies for the second transverse mode are determined by the
boundary value problem 
\bea 
\omega^2f+(G_c^2 f^\prime)^\prime&=&0 \nn
f^\prime|_{\sigma=0, p^+}=0 \;.  
\eea
{By comparison with Eq.~\ref{eq:eomtrans},} the spectrum of
  oscillation frequencies for this second transverse mode will differ from
  those for the ``manifest'' transverse mode {due only to the change in
the boundary conditions.}

\label{sec4}
\section{An interpolation between weak and strong coupling}
\label{ladder}
In this section we turn our attention to the field description 
of the dual to a free string stretched between two 1-branes on
the AdS boundary: the response of gauge fields
to a quark and antiquark constrained to move on two lines
parallel to the $z$-axis in the 't Hooft limit $N_c\to\infty$.
We should therefore try to sum all of the planar 
Feynman diagrams contributing to this process.
This is a daunting prospect, for which the lightcone worldsheet formalism
\cite{bardakcit,thorngauge,gudmundssontt} might be useful.
At weak coupling it suffices to sum only the ladder diagrams,
a tractable subset of all planar diagrams. Here we discuss the
ladder sum in both Feynman and lightcone gauge. The
corresponding Bethe-Salpeter equation that sums ladder diagrams 
is tractable at all coupling strengths. As already
noted in \cite{ericksonssz,klebanovmt} its solution at strong
coupling captures
many qualitative features of the known strong coupling limit
in the ${\cal N}=4$ case. This makes it an instructive model of
the interpolation between weak and strong coupling.
\subsection{Bethe-Salpeter equation for heavy sources on the lightcone}
In momentum space the ladder approximation to the Bethe-Salpeter
equation \cite{bethes} for a massive scalar quark-antiquark system, 
assuming the 't Hooft limit and 
with $P,p$ the total and relative momenta of the quarks, reads
\bea
&&\hskip-1in(m^2+(P/2 + p)^2)(m^2+(P/2-p)^2)\Psi_P(p)=\\
&&{g^2N_c\over2}\int {-id^4k\over(2\pi)^4k^2}
(P+2p-k)^\mu(-P+2p-k)^\nu N_{\mu\nu}(k)\Psi_P(p-k) \; ,
\eea
where $N_{\mu\nu}=\eta_{\mu\nu}$ in Feynman gauge and 
$N_{\mu\nu}=\eta_{\mu\nu}-k_\mu\eta_{\nu-}/k_--k_\nu\eta_{\mu-}/k_-$
in lightcone gauge. To adapt this equation to heavy external sources,
we take $m\to\infty$ and look for mass eigenstates of total mass
$M=2m-B$ with $B$, the binding energy, finite. Since $P^2=-M^2$,
we also neglect $p,k$ in comparison to $P$. Then the equation reduces to
\bea
\left({B^2\over4}-{(p\cdot P)^2\over M^2}\right)\Psi_P(p)=
-{g^2N_c\over2}\int {-id^4k\over(2\pi)^4k^2}
{P^\mu P^\nu N_{\mu\nu}(k)\over M^2}\Psi_P(p-k) \; .
\eea
To simplify even further we can work in the center of mass frame
$P=(M,{\vec 0})$:
\bea
\left({B^2\over4}-(p^{0})^2\right)\Psi_P(p)=
-{g^2N_c\over2}\int {-id^4k\over(2\pi)^4k^2}
N_{00}(k)\Psi_P(p-k) \; .
\eea
To describe static sources with separation ${\vec L}$ we just
put $\Psi_P=\phi(p^0)e^{i{\vec L}\cdot{\vec p}}$ and Fourier transform
$\phi(p^0)$ with respect to $t$ obtaining
\bea
\left({B^2\over4}+\partial_t^2\right)\psi(t)=
-{g^2N_c\over2}\int {-id^4k\over(2\pi)^4k^2}e^{i{\vec L}\cdot{\vec p}-ip^0t}
N_{00}(k)\psi(t) \; .
\eea
In Feynman gauge, $N_{00}=\eta_{00}=-1$ and the Fourier integral is elementary,
yielding, after Wick rotation $it\to t$, the Schr\"odinger 
equation analyzed in \cite{ericksonssz}:
\bea
\left[-\partial^2_t-{\lambda\over L^2+t^2}\right]g(t)=-{B^2\over4}g(t) \; ,
\label{onedsch}
\eea
where $\lambda=\alpha_s N_c/2\pi$.
In the language of the Wilson loop, this system corresponds
to a rectangular loop of width $L$ and length $T\to\infty$.
In the context of the ${\cal N}=4$ AdS/CFT correspondence 
the authors of \cite{ericksonssz} studied this approximation as a guide to
the physics of the interpolation between weak and 
strong coupling. Later it was shown in \cite{ericksonsz,grossd} that
for a circular Wilson loop the corresponding
``rainbow'' graph approximation
is actually exact in the limit $N_c\to\infty$, provided
scalar ladder rungs are also included with a strength such that
$\lambda$ doubles: $\lambda\to N_c\alpha_s/\pi$. This is not the
case for rectangular Wilson loops, but it is natural to 
include the scalar rungs when applying the interpolation to
the ${\cal N}=4$ case.

From the point of view of lightcone quantization, we interpret
the equation a little differently. We first consider sources free
to move on 1-branes parallel to the $z$ axis separated by a distance $L$.
This means putting $\Psi_P(p)=\phi(p^+,p^-)e^{i{\bfs L}\cdot{\bfs p}}$,
and Fourier transforming $\phi(p^+,p^-)$ in the variable $p^-$
with respect to lightcone time $x^+$. Recalling that $p^0=(p^++p^-)/\sqrt2$,
this leads to
\bea
\left({B^2\over4}-{1\over2}(p^++i\partial_+)^2\right)\psi(p^+,x^+)
=-{g^2N_c\over2}\int {-id^4k\over(2\pi)^4k^2}e^{i{\bfs L}\cdot
{\bfs k}-ik^-x^+}N_{00}(k)\psi(p^+-k^+,x^+) \; .
\eea
The integrations over $k^-$ and ${\bfs k}$
can be easily done for both the Feynman and lightcone gauge choices.
\bea
D(k^+,x^+) &\equiv& -\int {-id{\bfs k}dk^-\over(2\pi)^4k^2}
N_{00}e^{i{\bfs L}\cdot{\bfs k}-ik^-x^+} \nonumber\\
&=&{\theta(x^+k^+){\rm sgn}(k^+)\over8\pi^2\tau}e^{-k^+L^2/2\tau} \; , \qquad\qquad\qquad \mbox{ Feynman Gauge}\\
&=&{\theta(x^+k^+){\rm
    sgn}(k^+)\over16\pi^2\tau}\left[{L^2\over\tau^2}-{2\over
    k^+\tau}\right]e^{-k^+L^2/2\tau} \; , \;\;
\mbox{lightcone Gauge}
\eea
where $\tau\equiv ix^+$ will eventually be taken to be real.
The equation now takes the form
\bea
\left({B^2\over4}-{1\over2}(p^++i\partial_+)^2\right)\psi(p^+,x^+)
={g^2N_c\over2}\int dk^+ D(k^+,x^+)\psi(p^+-k^+,x^+) \; .
\eea
To see that this equation in Feynman gauge 
has identical content to (\ref{onedsch}),
put $\psi=e^{i(x^++a)p^+}f(x^+)$, which corresponds to putting
$z=a/\sqrt2$, to obtain
\bea
\left({B^2\over4}+{1\over2}\partial_+^2\right)f(x^+)
={g^2N_c\over2}\int dk^+ D(k^+,x^+)e^{-i(x^++a)k^+}f(x^+) \; .
\eea
Then for Feynman gauge we find
\bea
\hskip-.3in\int  dk^+ D(k^+,x^+)e^{-i(x^++a)k^+}&=&
{\theta(x^+)\over8\pi^2\tau}\int_0^\infty  dk^+e^{-k^+L^2/2\tau-i(x^++a)k^+} \nonumber\\
&-&{\theta(-x^+)\over8\pi^2\tau}
\int_{-\infty}^0 dk^+e^{-k^+L^2/2\tau-i(x^++a)k^+}\nonumber\\
&=&{1\over4\pi^2}{1\over L^2+a^2/2-2(x^{+}+a/2)^2} \; .
\eea
But since this solution fixes $z=a/\sqrt2$, $\sqrt{L^2+a^2/2}$ is
precisely the separation between the sources, so the change of
variables $t=i\sqrt2 (x^++a/2)$ reduces the lightcone Feynman gauge equation
to (\ref{onedsch}). For completeness we also work out
$D(k^+,x^+)$ for lightcone gauge, putting $a=0$ for simplicity.
\bea
\int dk^+ D(k^+,x^+)e^{-ix^+k^+}= 
{1\over4\pi^2}{L^2/2\tau^2\over L^2+2\tau^2}
&-&{\theta(x^+)\over8\pi^2\tau^2}\int_0^\infty {dk^+\over
k^+}e^{-k^+L^2/2\tau-ix^+k^+}\nonumber\\
&-&{\theta(-x^+)\over8\pi^2\tau^2}\int_0^\infty {dk^+\over
k^+}e^{k^+L^2/2\tau+ix^+k^+} \; .
\eea
This expression is problematic because the integrals diverge logarithmically
at $k^+=0$. Let's regulate the integrals by inserting a factor
$(k^+/\mu)^\alpha$, and use
\bea
\int_0^\infty {dk^+\over k^+}(k^+/\mu)^\alpha
e^{-k^+A}=\Gamma(\alpha)(\mu A)^{-\alpha}\to {1\over\alpha}+\Gamma^\prime(1)
-\ln(\mu A) \; .
\eea
Then one finds after some rearrangement and putting $t=\tau\sqrt2$,
\bea
\int dk^+ D(k^+,x^+)e^{-ix^+k^+} = {1\over4\pi^2}\left[{1\over t^2}
\ln{L^2+t^2\over L^2}-{1\over L^2+t^2}\right]
-{1\over4\pi^2t^2}\left[{1\over\alpha}+\Gamma^\prime(1)
-\ln{\mu L^2\over t\sqrt2}\right] \; .
\label{lckernel} 
\eea
The second term on the right is divergent as $\alpha\to0$ and violates
scale invariance and is singular as $t\to0$. Its presence indicates
that the naive ladder approximation in lightcone gauge is deficient
due to $k^+=0$ singularities.
On the other hand, we know from detailed one-loop studies that
in a complete gauge invariant 
computation these singularities are harmless, see for example
\cite{dipankarqt1,dipankarqt2}. The ladder diagrams are not gauge invariant
so it is not inconsistent for these singularities to cause
problems in them. In fact, different regulators of the $k^+=0$
singularities give different intermediate results. It is 
straightforward to show, for example, that the Mandelstam-Leibbrandt
(ML) principal value prescription \cite{mandelstam,leibbrandt}
applied to the Fourier transform of the lightcone gauge gluon
propagator gives precisely the first term on the right of
(\ref{lckernel}). That is, it interprets the divergent
second term as zero. If we use the ML prescription, then the 
lightcone gauge version of the potential term appearing in
(\ref{onedsch}) is
\be
V_{\rm LCML}(t)=-\lambda\left[{1\over t^2}
\ln{L^2+t^2\over L^2}-{1\over L^2+t^2}\right] \; .
\ee
In the weak coupling limit this potential can be replaced by a delta
function with coefficient
\be
\int dt V_{\rm LCML}(t) = -\lambda\int dt\left[{1\over t^2}
\ln{L^2+t^2\over L^2}-{1\over L^2+t^2}\right]
 = -\lambda\int dt{1\over L^2+t^2}=-\pi\lambda \; ,
\ee
which follows immediately from integrating the first term in square
brackets by parts. We see that this agrees with the integral 
of the analogous potential from Feynman gauge. Thus the ML
treatment of lightcone gauge ladder diagrams 
agrees in weak coupling with Feynman gauge.
In the remainder of this section we further analyze the ladder
model using Feynman gauge, but the lightcone gauge a la ML should give
similar results.

\subsection{Ladder diagram model of the spectrum}

Ref. \cite{ericksonssz} showed that in the weak coupling limit the
Schr\"odinger equation (\ref{onedsch}) has a single bound state with
 $B=\pi\lambda/L$ which is just the Coulomb interaction energy
$-N_cg^2/4\pi L=-N_c\alpha_s/2L$ in pure QCD (or twice this
for ${\cal N}=4$) between heavy quark and antiquark
in the fundamental representation. They also noted that in the
strong coupling limit the Schr\"odinger equation is well described by its
harmonic oscillator approximation
\bea
\left[-\partial^2_t-{\lambda\over L^2}+{\lambda\over L^4}t^2\right]g(t)
=-{B^2\over4}g(t) \; ,
\eea
whose natural frequency is $\omega_0=\sqrt{2\lambda}/L^2$, so
\bea
B_N=2\sqrt{{\lambda\over L^2}-{\sqrt{2\lambda}\over2L^2}-N{\sqrt{2\lambda}
\over L^2}}
\simeq {2\sqrt{\lambda}\over L}-{\sqrt2\over2L}-N{\sqrt2\over L} \; .
\eea

In the case of lightcone gauge the potential $V_{LCML}(t)$
has minima away from $t=0$ at $t=\pm x_0L$ where $x_0$ satisfies
\bea
\ln(1+x_0^2)={x_0^2(1+2x_0^2)\over(1+x_0^2)}
\eea
for which $x_0\simeq 1.47$. In this case the harmonic approximation
reads
\be
V_{LCML}(t )=  -{\lambda\over L^2}{x_0^2\over(1+x_0^2)^2}+{2\lambda\over L^4}
{x_0^2-1\over (x_0^2+1)^3}(t\pm x_0 L)^2
\simeq -0.216{\lambda\over L^2}+.0735{\lambda\over L^4}(t\pm x_0 L)^2 \; .
\ee
So besides a difference in the numerical coefficients, the lightcone
gauge case shows two families of nearly degenerate oscillator levels
at strong coupling.

We close this subsection with an analysis of the
Feynman gauge ladder model to 
find the coupling at which, in this model, the first discrete levels
peel off the continuum. In terms of the Schr\"odinger equation
(\ref{onedsch}) the first discrete level is the first excited
state which is present only for sufficiently large coupling.
To do this let us first simplify the equation by
defining $x=t/L$ and $b=BL/2$:
\bea
\left[-\partial^2_x-{\lambda\over 1+x^2}\right]g=-{b^2}g \; .
\eea
When $x\gg1$ the solution to this equation is just the Kelvin
function $g_\infty(x)=\sqrt{x}K_\nu(bx)$, with $\nu=\sqrt{1/4-\lambda}$.
We want to find the smallest $\lambda$ for which a discrete level
forms just below threshold, i.e. whose eigenvalue $b^2\ll1$.
In this case there is a region of $x$, $1\ll x \ll 1/b$
for which the small argument approximation of $K_\nu$ is valid:
\bea
g_\infty(x)=\sqrt{x}K_\nu(bx)\simeq \sqrt{x}
{\pi\over2\sin\pi\nu}\left[{1\over\Gamma(1-\nu)}
\left({bx\over2}\right)^{-\nu}-{1\over\Gamma(1+\nu)}
\left({bx\over2}\right)^{\nu}\right] \; .
\label{smallargk}
\eea 
Note that in the range $0<\lambda<1/4$ for which $\nu$ is
real and positive, the second term in this approximation
can be neglected. 

On the other hand, in the region 
$0<x\ll\sqrt{\lambda}/b$, the solution is well-approximated by the
solution $g_0$ of the Schr\"odinger equation with $b=0$
\bea
\left[-\partial^2_x-{\lambda\over 1+x^2}\right]g_0=0 \; .
\eea
Since we are searching for the first excited state, {$g$ and $g_0$}
must vanish at $x=0$. The solution of this last equation is
\bea
g_0(x)=\sqrt{1+x^2}(A_1P^1_{\nu-1/2}(ix)+A_2Q^1_{\nu-1/2}(ix)) \; ,
\eea
where $P^1_{\nu-1/2},Q^1_{\nu-1/2}$ are the Legendre functions. 
 It follows from the differential equation that in the region
$x\gg1$ it has the behavior
\bea
g_0\simeq A\sqrt{x}(x^{\nu}+C(\lambda)x^{-\nu}) \; .
\eea
If $C(\lambda)$ is finite and $\lambda<1/4$ the second term of this
asymptotic behavior is negligible. And in the region of overlapping
validity matching $g_0$ and $g_\infty$ is impossible.

By investigating the properties of Legendre functions one
finds that the condition $g_0(0)=0$ implies:
\bea
 C(\lambda)=&&{2^{-2\nu}\Gamma(-\nu)\Gamma(\nu-1/2)\cos\pi(\nu/2+1/4)
\over\Gamma(\nu)\Gamma(-\nu-1/2)\cos\pi(-\nu/2+1/4) 
} \nonumber \\= & -& {2^{-2\nu}\Gamma(1-\nu)\Gamma(3/2+\nu)\sin\pi(1/4-\nu/2)
\over\Gamma(1+\nu)\Gamma(3/2-\nu)\sin\pi(1/4+\nu/2) 
} \; .
\eea
This result applies not only to the first excited level but to
all odd parity levels. 
Comparing to (\ref{smallargk}) we read off the matching condition
valid whenever $1\ll x\ll 1/b, \sqrt{\lambda}/b$:
\bea
\left({b\over2}\right)^{-2\nu}=
{2^{-2\nu}\Gamma(1-\nu)^2\Gamma(3/2+\nu)\sin\pi(1/4-\nu/2)
\over\Gamma(1+\nu)^2\Gamma(3/2-\nu)\sin\pi(1/4+\nu/2)
} \; .
\eea
If $\sqrt{\lambda}/b\gg1$, there is an overlapping region where
$g$ is simultaneously well approximated by both $g_0$ and
$g_\infty$. We see explicitly that the right side 
is finite in the range $0<\lambda<1/4$ 
so we conclude that there are no odd parity discrete levels with $b\ll1$
for $\lambda<1/4$. Since the first excited discrete level must be odd parity
and since the only discrete level for weak coupling is the
ground state,
it follows from continuity
that there are no excited discrete levels at all for these $\lambda$.
Thus the critical $\lambda$ we
are seeking must satisfy $\lambda_c\geq1/4$. 

Indeed it is exactly equal to $1/4$.
For $\lambda>1/4$, $\nu=i\sqrt{\lambda-1/4}$ and both terms in the
asymptotic approximations of $g_\infty$ and $g_0$ must be kept
and there is sufficient flexibility to match the two 
asymptotic behaviors in the region of common validity.
Indeed, it is easy to see that when $\nu$ is imaginary and
fixed there is actually an infinite accumulation of 
levels approaching threshold. Let $b_1\ll1$ be an eigenvalue
solving the matching condition. Then $b_1e^{-\pi n/|\nu|}$
for $n$ an integer also obeys the matching condition.
For all $n>0$ these new levels are closer to threshold than $b_1$
and hence all are valid eigenvalues.

For even parity states, we impose vanishing derivative $g_0^\prime(0)=0$
and following the same steps arrive at
\bea
C^{even}(\lambda)
&=&-{2^{-2\nu}\Gamma(1-\nu)\Gamma(3/2+\nu)\sin\pi(1/4+\nu/2)
\over\Gamma(1+\nu)\Gamma(3/2-\nu)\sin\pi(1/4-\nu/2)
} \; .
\eea
Putting $\nu=i|\nu|$ we find
\bea
{\sin\pi(1/4+\nu/2)
\over\sin\pi(1/4-\nu/2)
}=i{1-ie^{-|\nu|}\over1+ie^{-|\nu|}},\qquad
{\sin\pi(1/4-\nu/2)
\over\sin\pi(1/4+\nu/2)
}=-i{1+ie^{-|\nu|}\over1-ie^{-|\nu|}} \; .
\eea
At strong coupling these two factors differ by a sign. Thus the even levels
near threshold are uniformly interleaved between the 
odd levels at strong coupling: $E_n\simeq -b_1e^{-\pi n/2|\nu|}$,
in agreement with a slightly different argument in \cite{klebanovmt}.

We conclude that an infinite number of discrete levels
appear for any $\lambda>\lambda_c=1/4$. Referring back to the relation
between $\lambda$ and $\alpha_s$, we find the new states
coming in for
\bea
{N_c\alpha_s\over\pi}>{1\over2} \; .
\eea
For $N_c=3$ this gives $\alpha^c_s=\pi/6\simeq 0.523$, slightly bigger than
1/2. Though not tiny, this value of $\alpha_s$ is small enough to hope that
radiative corrections to the ladder approximation may be
under enough control near the critical
value to allow us to assess whether a similar
transition appears in large $N_c$ QCD itself. In that case of course the
coupling depends on the separation $\alpha_s(L)$, and we
should instead speak of a critical separation, above which
extra valence gluons may start playing a decisive role
in the confining gluonic flux tube \cite{greensitet}.
\subsection{A one gluon truncation of the ladder model}
Some insight into the physics of the discrete levels in the ladder
approximation can be gleaned by considering the intermediate states
included in the sum. Although ladders are an iteration of single
gluon exchange, there are unitarity cuts through them that include
arbitrary numbers of gluons. Because of the restriction
to ladder diagrams these gluons do not interact with
each other except for a veto for one crossing 
another. This will change with the inclusion
of other planar diagrams. We suggest that the discrete levels
are a reflection of these states with high numbers of gluons.

As evidence for this we study a truncation of the ladder sum that
removes the contribution of intermediate states with more than
one gluon. To do this we first present the Bethe-Salpeter equation in
the form of an integral equation:
\bea
\psi(T,U)=\lambda\int dt du\ \theta(T-t)\theta(U-u){1\over (t-u)^2+L^2}
\psi(t,u) \; .
\eea 
With $t<T$, $t$ and $T$ mark times for two sequential gluon emission from
(absorption by) one source whereas $u$ and $U$, with $u<U$, mark times for
the absorption by (emission from) the other.  However, multi-gluon states are
still included because the integral includes regions with $t>U$ or $u>T$. We
can veto this possibility by including two more step functions in the
integrand:
\bea
\psi(T,U)=\lambda\int dt du\ \theta(T-t)\theta(T-u)
\theta(U-u)\theta(U-t){1\over (t-u)^2+L^2}\psi(t,u) \; .
\eea 
Differentiating this modified equation with respect to $T$ and $U$
leads to an integro-differential equation:
\bea
{\partial^2\psi\over\partial T\partial U}=
\lambda\delta(T-U)\int du \theta(T-u)\theta(U-u)
{1\over (u-U)^2+L^2}(\psi(U,u)+\psi(u,U)) \; .
\eea
Energy eigenstates correspond to the ansatz 
$\psi(T,U)=e^{-E(T+U)/2}f(T-U)$, resulting in the equation
\bea
\left({E^2\over4}-\partial_T^2\right)f(T)=\lambda\delta(T)\int_{-\infty}^0
du{1\over u^2+L^2}(f(u)+f(-u))e^{-Eu/2} \; ,
\eea
which tacitly assumes we are searching for a bound state $E<0$.
Clearly the Fourier transform of $f$ is given by
\bea
{\tilde f}(\omega)&=&\int dT e^{i\omega T}f(T)={\lambda C\over\omega^2+E^2/4}\\
C&=&\int_{-\infty}^0
du{1\over u^2+L^2}(f(u)+f(-u))e^{-Eu/2} \; .
\eea
Then
\bea
{f}(T)&=&\int {d\omega\over2\pi} e^{-i\omega T}{\tilde f}(\omega)
=C{\lambda\over|E|}e^{-|ET|/2} \; .
\eea
Plugging this into the expression for $C$ leads to the eigenvalue
equation
\bea
{1\over2\lambda}={1\over|E|}\int_0^\infty du{1\over u^2+L^2}e^{-|E|u}
={1\over|E|L}\int_0^\infty du{1\over u^2+1}e^{-|E|Lu}
\simeq\cases{{\pi\over2|E|L}& for $|E|L\ll1$\cr
{1\over L^2E^2}& for $|E|L\gg1$\cr}\; .
\eea
Clearly the right side is a monotonically decreasing function of
$|E|$ so there is exactly one solution for any $\lambda>0$.
The weak and strong coupling limits of the energy eigenvalue
are $E=-\pi\lambda/L$, $E=-\sqrt{2\lambda}/L$ respectively.
This one gluon truncation of the Bethe-Salpeter equation then
loses the discrete levels but still produces the $\lambda\to\sqrt\lambda$
replacement for weak to strong coupling. This exercise shows
the importance of multi gluon states in generating the discrete
levels that populate the gap.

\section{Discussion}

We have examined the spectrum of the string/flux tube in strong and weak
coupling respectively to understand the qualitative features of the AdS/CFT
correspondence. At first it is surprising that in the strong coupling limit
of a conformal theory with a pure AdS background and no confinement, one has
a discrete spectrum. However as explained in \cite{klebanovmt} this can be
understood in weak coupling as a consequence of the large $N_c$ limit. This
paper goes on to compare the results for this configuration using both the
lightcone gauge and temporal gauge on the strong coupling side and
lightcone and Feynman gauge in weak coupling.  In strong coupling to
quadratic order we find the equations of motion for both the transverse
and longitudinal modes with frequencies 
inversely proportional to the separation of
the sources $1/L$. 

The ladder approximation we used to analyze
the field side of the duality is obviously inadequate for large
`t Hooft coupling. Nonetheless, it is  encouraging that its strong coupling
limit is qualitatively similar to the exact AdS/CFT results. The
lowest energy eigenvalue has the $\sqrt{\lambda}/L$ coupling dependence 
in both cases although the dimensionless coefficients disagree for
obvious reasons. Further in both cases the gap between the lowest
energy and the continuum is populated with many discrete levels.
As shown in \cite{klebanovmt} there is also an exact match 
at strong coupling of the density of near threshold states
of the ladder model and the exact strong coupling results
given by the AdS/CFT correspondence.
The big qualitative difference on that score is the density of
states near the ground state for strong coupling. 
The AdS/CFT correspondence shows that these levels are those of a string,
namely those of an infinite number of oscillators with frequencies $\omega_n$,
$n=1,2\cdots$.
In contrast, the sum of ladder diagrams in Feynman gauge
implies discrete levels
of a single harmonic oscillator. Presumably, more complicated planar diagrams
which include interactions between the many
gluons in the intermediate states will remove this discrepancy.

Finally recall that the basic analysis on the string side was formulated for
a more general metric which can describe QCD like models of confining
backgrounds.  If the background approaches $AdS_5$ in the UV, the same
spectral analysis will hold for small L. Of course here a discrete spectrum
is required by confinement.  It is interesting however to consider the
effects of confinement and/or a running coupling $\lambda(L)$ in this limit
and to try to disentangle these effects from the large $N_c$ effects that are
responsible for the discrete spectrum in the conformal limit.  Indeed lattice
investigations for pure Yang-Mills theory (or quarkless QCD) have revealed a
rather intricate spectrum~\cite{Juge:2002br} for the small L stretched string
(or what is called gluelumps~\cite{Bali:2003jq}), which present a challenge
to the understanding of short strings in an approximately conformal
background.  These results are not expected to be affected dramatically by
taking the large {$N_c$} limit.  Thus a search for states corresponding to
the radion modes and/or fermionic degrees of freedom should be encouraged.
This could help substantially to guide the construction of more realistic
models of the QCD string dual to pure Yang-Mills theory.

\vskip12pt

\noindent\underline{Acknowledgments:}
We thank the organizers of the 
Fall 2004 Program on {\it QCD
and String Theory} at KITP, Santa Barbara, CA, 
for the stimulating environment in which this work began.
Also CBT thanks I. Klebanov and J. Maldacena for valuable
discussions. 
This work was supported in part by the Department of Energy
under Grants No.  DE-FG02-91ER40676, Task-T,  DE-FG02-97ER-41029 and
DE-FG02-91ER40688, Task-A.


\begin{thebibliography}{1}
\bibitem{thooftlargen}
G. 't Hooft, {\sl Nucl. Phys.} {\bf B72} (1974) 461.
\bibitem{maldacena}
J.~Maldacena,
Adv.\ Theor.\ Math.\ Phys.\ {\bf 2}, 231 (1998),
hep-th/9711200.   S.S.~Gubser, I.R.~Klebanov and A.M.~Polyakov,
Phys.\ Lett.\ {\bf B428}, 105 (1998),
hep-th/9802109.
E.~Witten,
Adv.\ Theor.\ Math.\ Phys.\ {\bf 2}, 253 (1998),
hep-th/9802150.
\bibitem{bmn}
  D.~Berenstein, J.~M.~Maldacena and H.~Nastase,
  JHEP {\bf 0204} (2002) 013
  [arXiv:hep-th/0202021].

\bibitem{maldacenaqqbar}
J. Maldacena, 
Phys. Rev. Lett. {\bf 80}, 4859 (1998), {hep-th/9803002};
S.-J. Rey and J. Yee,
{hep-th/9803001}.



\bibitem{callang}
C.~G.~Callan and A.~G\"uijosa,
Nucl.\ Phys.\ B {\bf 565} (2000) 157
[arXiv:hep-th/9906153].
\bibitem{Bak}
  D.~Bak and S.~J.~Rey,
  Nucl.\ Phys.\ B {\bf 572}, 151 (2000)
  [arXiv:hep-th/9902101].
\bibitem{klebanovmt}
I.~R.~Klebanov, J.~Maldacena and C.~B.~Thorn,
  ``Dynamics of Flux Tubes in Large N Gauge Theories,''
  arXiv:hep-th/0602255.
\bibitem{bardakcit}
  K.~Bardakci and C.~B.~Thorn,
  Nucl.\ Phys.\ B {\bf 626} (2002) 287
  [arXiv:hep-th/0110301].
\bibitem{thorngauge}
  C.~B.~Thorn,
  Nucl.\ Phys.\ B {\bf 637} (2002) 272
  [Erratum-ibid.\ B {\bf 648} (2003) 457]
  [arXiv:hep-th/0203167].
\bibitem{gudmundssontt}
  S.~Gudmundsson, C.~B.~Thorn and T.~A.~Tran,
  Nucl.\ Phys.\ B {\bf 649} (2003) 3
  [arXiv:hep-th/0209102].
\bibitem{metsaevtt}
  R.~R.~Metsaev, C.~B.~Thorn and A.~A.~Tseytlin,
  Nucl.\ Phys.\ B {\bf 596} (2001) 151
  [arXiv:hep-th/0009171].
\bibitem{ericksonssz}
J.~K.~Erickson, G.~W.~Semenoff, R.~J.~Szabo and K.~Zarembo,
Phys.\ Rev.\ D {\bf 61} (2000) 105006
[arXiv:hep-th/9911088]. 
\bibitem{greensitet}
  J.~Greensite and C.~B.~Thorn,
  JHEP {\bf 0202} (2002) 014
  [arXiv:hep-ph/0112326].
\bibitem{btt}
R.~C.~Brower, C-I~Tan,  and E.~Thompson,
Int. J. Mod. Phys. A {\bf20} (2005) 4508
[arXiv:hep-th/0503223].


\bibitem{rozowskyt}
  J.~S.~Rozowsky and C.~B.~Thorn,
  Phys.\ Rev.\ D {\bf 60} (1999) 045001
  [arXiv:hep-th/9902145].


\bibitem{Arvis:1983fp}
  J.~F.~Arvis,
  Phys.\ Lett.\ B {\bf 127}, 106 (1983).






\bibitem{bethes}
E.~E.~Salpeter and H.~A.~Bethe, {\sl Phys. Rev.} {\bf 84} (1951) 1232.

\bibitem{ericksonsz}
J.~K.~Erickson, G.~W.~Semenoff and K.~Zarembo,
Nucl.\ Phys.\ B {\bf 582} (2000) 155
[arXiv:hep-th/0003055]. 
\bibitem{grossd}
N.~Drukker and D.~J.~Gross,
  J.\ Math.\ Phys.\  {\bf 42} (2001) 2896
  [arXiv:hep-th/0010274].


\bibitem{dipankarqt1}
  D.~Chakrabarti, J.~Qiu and C.~B.~Thorn,
  Phys.\ Rev.\ D {\bf 72} (2005) 065022
  [arXiv:hep-th/0507280].
\bibitem{dipankarqt2}
  D.~Chakrabarti, J.~Qiu and C.~B.~Thorn,
  ``Scattering of glue by glue on the lightcone worldsheet. II: Helicity
  conserving amplitudes,''
  arXiv:hep-th/0602026.
\bibitem{mandelstam}
  S.~Mandelstam,
  Nucl.\ Phys.\ B {\bf 213} (1983) 149.
\bibitem{leibbrandt}
  G.~Leibbrandt,
  Phys.\ Rev.\ D {\bf 29} (1984) 1699.
\bibitem{Juge:2002br}
  K.~J.~Juge, J.~Kuti and C.~Morningstar,
  Phys.\ Rev.\ Lett.\  {\bf 90}, 161601 (2003)
  [arXiv:hep-lat/0207004].
 \bibitem{Bali:2003jq}
  G.~S.~Bali and A.~Pineda,
  Phys.\ Rev.\ D {\bf 69}, 094001 (2004)
  [arXiv:hep-ph/0310130].
\end{thebibliography}
\end{document}